\documentclass[twocolumn,showpacs,preprintnumbers,amsmath,amssymb,superscriptaddress, prb]{revtex4-1}
\usepackage{amssymb}
\usepackage{graphicx}
\usepackage{dcolumn}
\usepackage{bm}
\usepackage{times}
\usepackage{hyperref}
\usepackage{color}
\usepackage{todonotes}
\usepackage{textcomp}
\newcommand{\mss}{Mn$_2$SiS$_4$}
\newcommand{\msse}{Mn$_2$SiSe$_4$}
\newcommand{\mssse}{Mn$_2$SiS$_{4-x}$Se$_x$}%

\begin{document}
%
\title{Antiferromagnetism and the emergence of frustration in saw-tooth lattice chalcogenide olivines Mn$_2$SiS$_{4-x}$Se$_x$ ($x$ = 0 $\textendash$ 4)}
%
%
\author{H. Nhalil} 
\affiliation{Department of Chemistry and Biochemistry, University of Oklahoma, 101 Stephenson Parkway, Norman, OK 73019, USA}
\author{R. Baral} 
\affiliation{Department of Physics, 500 W University Ave, University of Texas at El Paso, El Paso, TX 79968, USA}
\author{B. O. Khamala}
\affiliation{Department of Physics, 500 W University Ave, University of Texas at El Paso, El Paso, TX 79968, USA}
\author{A. Cosio} 
\affiliation{Department of Physics, 500 W University Ave, University of Texas at El Paso, El Paso, TX 79968, USA}
\author{S. R. Singamaneni} 
\affiliation{Department of Physics, 500 W University Ave, University of Texas at El Paso, El Paso, TX 79968, USA}
\author{M. Fitta} 
\affiliation{The Henryk Niewodnicza\'{n}ski Institute of Nuclear Physics -PAN, Department of Magnetic Materials and Nanostructures, ul. Radzikowskiego 152, 31-342 Krak\'{o}w, Poland}
\author{D. Antonio}
\affiliation{Idaho National Laboratory, Idaho Falls, ID 83415, USA}
\author{K. Gofryk}
\affiliation{Idaho National Laboratory, Idaho Falls, ID 83415, USA}
\author{R. R. Zope} 
\affiliation{Department of Physics, 500 W University Ave, University of Texas at El Paso, El Paso, TX 79968, USA}
\author{T. Baruah} 
\affiliation{Department of Physics, 500 W University Ave, University of Texas at El Paso, El Paso, TX 79968, USA}
\author{B. Saparov} 
\email{saparov@ou.edu}
\affiliation{Department of Chemistry and Biochemistry, University of Oklahoma, 101 Stephenson Parkway, Norman, OK 73019, USA}
\author{H. S. Nair} 
\email{hnair@utep.edu}
\affiliation{Department of Physics, 500 W University Ave, University of Texas at El Paso, El Paso, TX 79968, USA}
\date{\today} 

\begin{abstract}
The magnetism in the saw-tooth lattice of Mn in the 
olivine chalcogenides, Mn$_2$SiS$_{4-x}$Se$_x$ ($x$ = 1$\textendash$4) 
is studied in detail by analyzing their magnetization, specific heat
and thermal conductivity properties and complemented with 
density functional theory calculations.
The air-stable chalcogenides are antiferromagnets and 
show a linear trend in the transition temperature, $T_N$ as 
a function of Se-content ($x$) which shows a decrease 
from $T_N \approx$ 86~K for {\mss} to 66~K for {\msse}.
Additional new magnetic anomalies are revealed at low temperatures for
all the compositions. Magnetization irreversibilities are also observed
as a function of $x$. The specific heat and the magnetic
entropy indicate the presence of short-range spin fluctuations 
in Mn$_2$SiS$_{4-x}$Se$_x$.
A spin-flop antiferromagnetic phase transition in the presence of
applied magnetic field is present in Mn$_2$SiS$_{4-x}$Se$_x$, 
where the critical field for the spin flop increases from $x$ = 0 towards 4
in a non-linear fashion. 
Density functional theory calculations show that an overall
antiferromagnetic structure with ferromagnetic coupling of the
spins in the $ab$-plane minimizes the total energy.
The band structures calculated for \mss\ and \msse\ reveal
features near the band edges similar to those reported for
Fe-based olivines suggested as thermoelectrics; however the experimentally determined thermal transport data do not support superior thermoelectric features.
The transition from long-range magnetic order in \mss\ to short-range order and spin fluctuations in \msse\ is explained
using the variation of the Mn-Mn distances in the triangle units
that constitutes the saw-tooth lattice upon progressive replacement
of sulphur with selenium.
\end{abstract}
\maketitle
%

\section{introduction}
\indent 
Complex magnetic excitations from frustrated lattices of magnetic
atoms is an attractive topic in quantum correlated systems.
The saw-tooth antiferromagnetic chain has a frustrated topology
of corner-sharing triangles of spins where the ground state of the 
spin-half saw-tooth chain is understood exactly \cite{kubo1993k, sen1996quantum, nakamura1996t}.
Variety of ground states are predicted for the saw-tooth lattice
depending on the ratio of the exchange interaction strengths between
the base-base and the base-vertex pairs \cite{blundell2003quantum, ohanyan2009v,bellucci2010lattice,hao2011destruction,chandra2004vr}.
The saw-tooth systems attain importance in connection with the 
zero energy flat-band modes similar to the case of Kagome lattices \cite{zhitomirsky2004me,zhitomirsky2005me} and are valuable as potential materials for magnonics \cite{wang2018topological}.
Experimental studies on saw-tooth lattices are
limited in number; some examples are the
delafossites, olivine and germanates \cite{cava1993rj,le2005bacq,lau2006gc,white2012coupling}.
In this connection, chalcogenide olivines have received
less attention regarding the magnetism and magnetic 
excitations arising from their underlying saw-tooth
lattice. 
The $A_2BX_4$ ($A$ = Mn, Fe, Ni; $B$ = Si, Ge; $X$ = S, Se, Te, O)
olivines, where the $A$ atoms form a saw-tooth lattice, are well-known semiconducting magnetic compounds which find 
applications in optoelectronics and magentic devices
\cite{fredrick_solution_2013,furdyna_diluted_1988}.
They have been recently computationally projected as 
suitable thermoelectric candidates owing to peculiar band structure
features \cite{gudelli_predicted_2015}.
They crystallize in orthorhombic $Pnma$ space group and have relatively 
small tetrahedral ions ($B$) and large octahedral ions ($A$). 
Olivines have a spinel-like structure but uses one quadravalent and two 
divalent cations $A_2^{2+}B^{4+}X_4$ instead of 
two trivalent and one divalent cations. The $A$ sites consist of the two crystallographically independent sites ($4a$ and $4c$ sites) 
and form a triangle-based saw-tooth chain structure through the edge-sharing 
bonds along the $b$-axis\cite{hagemann_geometric_2000}.
Due to this structural feature, the $A$-site lattice is 
geometrically magnetically frustrated when it is occupied by magnetic ions 
\cite{hagemann_geometric_2000}.
The end-compounds of \mssse\ $\textendash$ \mss\ and \msse\ $\textendash$ order 
antiferromagnetically below their Ne\'{e}l temperature, $T_N \approx$ 
83~K and 66~K, respectively
\cite{ohgushi_anomalous_2005,lamarche_possible_1994,jobic_structure_1995,bodenan_low-temperature_1996}.
\mss\ belongs to the class of anisotropic uniaxial antiferromagnets
but with anomalous magnetic features near the
spin-flop transition \cite{ohgushi_anomalous_2005}.
A weak ferromagnetic interaction exists in a narrow temperature window between 83~K and 86~K, while displaying uniaxial anisotropy with the $b$-direction 
as the easy axis. The origin of weak ferromagnetism (WF) and the unusual temperature
dependence of spin-flop critical field is unclear in olivines despite the microscopic origin of WF which is supported by neutron scattering experiments \cite{lamarche_neutron_1994}. At 4.2~K a collinear ferromagnetic arrangement of the
Mn spins at the two distinct crystallographic positions, $4a$ (a site with inversion) and $4c$ (mirror), was observed along the $b$ axis. As the temperature increases to 
83~K, the orientation of the $4a$ spins turns in the $ab$ plane.
At 83~K, both the $4a$ and the $4c$ spins reorient along the $a$ axis but with 
some canting in the $ac$ plane. It is in the very small temperature range of 83$\textendash$86~K, spins at two different crystallographic positions
display weak ferromagnetism.
The paramagnetic to antiferromagnetic transition has been
identified as belonging to the Heisenberg universality class and the weak ferromagnetic transition as first order with a latent heat $\approx$ 0.01~J/mol 
\cite{junod_specific_1995}. A very low value of magnetic entropy, about 5$\%$
of $R$ln(2$S$ + 1), is found to be released at the antiferromagnetic transition, indicating that the spin entropy is not completely removed at the $T_N$. 
Experimental support for the short-range spin fluctuations come from the fact that purely magnetic intensity was observed in neutron diffraction data up to 140~K \cite{lamarche_neutron_1994}.\\
\indent
On the other end of the composition series of \mssse\ is the 
case of Mn$_2$SiSe$_4$, which has the magnetic easy axis along 
crystallographic $c$-direction of the orthorhombic cell
\cite{bodenan_low-temperature_1996}. 
In the case of \msse, the average magnetic structure
remains in a configuration intermediate to a ferrimagnet and
an antiferromagent for most of the $T < T_N$ region. 
Though both \mss\ and \msse\ are reported to
show similar magnitude of magnetization, \msse\
displays pronounced field and temperature cycling
dependencies in magnetic susceptibility \cite{bodenan_low-temperature_1996}.
The temperature range spanned by the magnetization maximum 
(between 66~K and 17~K,  almost 50~K) is much wider compared to 
that of \mss\ (between 86~K and 83~K, approximately 3~K)
\cite{jobic_structure_1995,bodenan_low-temperature_1996}.
The broadness of the transition in magnetic susceptibility
of \msse\ and the hysteresis-like effects already
suggests strongly competing interactions leading
to a frustrated magnetic state.
An interesting aspect of the olivines that has recently
received attention is related to thermoelectricity. 
Quasi-flat band edges near the valence and conduction bands
were predicted using density functional theory calculations
in the case of Fe$_2$Ga$Ch_4$ ($Ch$ = S, Se, Te) \cite{gudelli_predicted_2015}.
This theoretical investigation was preceded by experiments
that showed nano-structured Fe$_2$GeS$_4$ is 
a photovoltaic material \cite{park2015highly, fredrick_solution_2013}. 
Experimental studies have shown that Fe$_2$SiS$_4$
and Fe$_2$GeS$_4$ possess significant thermopower \cite{platt_copper_2010}.
\\
\indent 
In the present paper we undertake a detailed experimental 
study of magnetism in \mssse. Our research is motivated
by the prospect of understanding the role of magnetic frustration in
the saw-tooth lattice of Mn in the series of olivines as the
transition metal environment is altered from sulfur-rich to selenium-rich. 
A detailed magnetic and thermal property investigation of 
Mn$_2$SiS$_{4-x}$Se$_x$ ($x$ = 0, 1, 2, 3, 4) solid solutions
is undertaken and is complemented with density functional
theory calculations.
\begin{figure}[!t]
	\centering\includegraphics[scale=0.06]{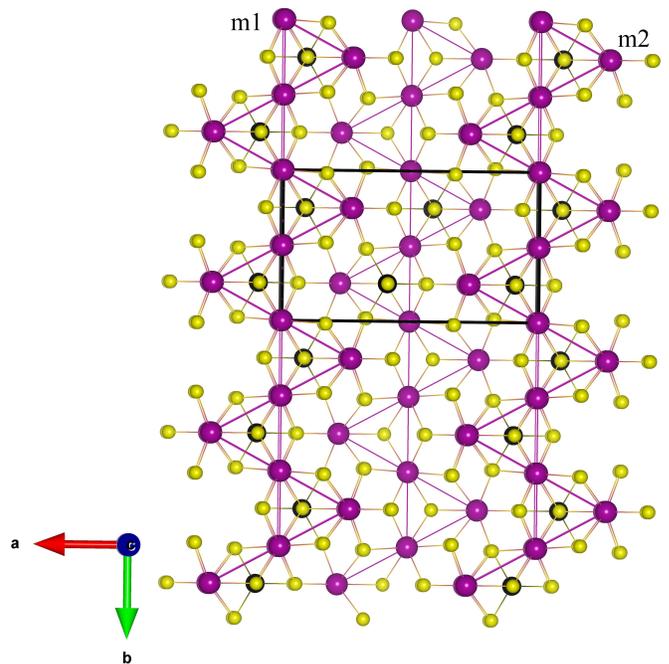}
	\caption{\label{fig_str} (color online) 
        A schematic of the olivine structure of {\mss}. The unit cell is
        outlined in black solid line. The purple spheres are Mn. The saw-tooth lattice formed by Mn$^{2+}$ at the two crystallographically-inequivalent
		sites, m$_1$ and m$_2$ is shown. The line of m$_1$ atoms forms
		along the $b$ direction. In the figure, the black spheres are Si, and the 
		yellow are S. The figure was created using VESTA\cite{momma_vesta_2011}.}
\end{figure}

\section{Methods}
\subsection{Experimental techniques}
\indent
Elemental Mn, Si, S and Se (99.99$\%$, Aldrich) were used as reactants
to synthesize \mssse, $x$ = 0$\textendash$4. Stoichiometric amount of these elements 
were weighed and mixed properly using a mortar and pestle before pelleting 
and loading into a 10~mm diameter quartz ampule in a N$_2$-filled glove box. 
The quartz tubes were flame-sealed under a dynamic vacuum with pressure 
less than 10$^{-3}$ mTorr. The reaction mixtures were heated at 1000$^{\circ}$C
for 24~hours then cooled to room temperature at a rate of 100$^{\circ}$C/h. 
To improve the phase purity and crystallinity, samples were re-ground, pelleted 
and annealed under identical conditions as necessary. 
Room temperature powder X-ray diffraction (PXRD) measurements 
were performed on a Rigaku MiniFlex600 instrument with a D/tex detector 
using a Ni-filtered Cu-$K\alpha$ radiation 
($\lambda_1$: 1.540562 $\AA$; $\lambda_2$: 1.544398 $\AA$). 
X-ray data collection experiments were performed at room temperature 
in the 10-70$^{\circ}$ (2$\theta$) range, with a step size of 0.02$^{\circ}$. 
Data analysis was performed using Rigaku PDXL software package. 
The collected data were fitted using the decomposition method 
(Pawley fitting) embedded in the PDXL package.
For air stability studies, powder samples of all three compositions were 
left in ambient air for a period of 6 weeks. 
PXRD measurements were regularly performed during this 
period using the conditions described above.
The specific heat, $C_p(T)$, of the samples were measured
using the heat pulse method in a commercial VersaLab,
Physical Property Measurement System from Quantum Design.
Tiny pellets of {\mssse} of mass approximately 2-3~mg were
used for the measurements. The sample was attached to the
calorimeter puck using N Apiezon grease. The $C_p(T)$ was
measured in the temperature range 50~K$\textendash$300~K
under  0~T and 3~T.
The temperature and field-dependent magnetization measurements
were performed in a SQUID Magnetic Property Measurement System. 
DC magnetization was measured
in the temperature range 2$\textendash$300~K and isothermal magnetization
at 2~K in the range $\textendash$7~T to 7~T. 
The thermal conductivity was measured using the 
TTO option in a commercial DynaCool-9, Physical 
Property Measurement System from Quantum Design. \\
\begin{figure}[!b]
	\centering
	\includegraphics[scale=0.54]{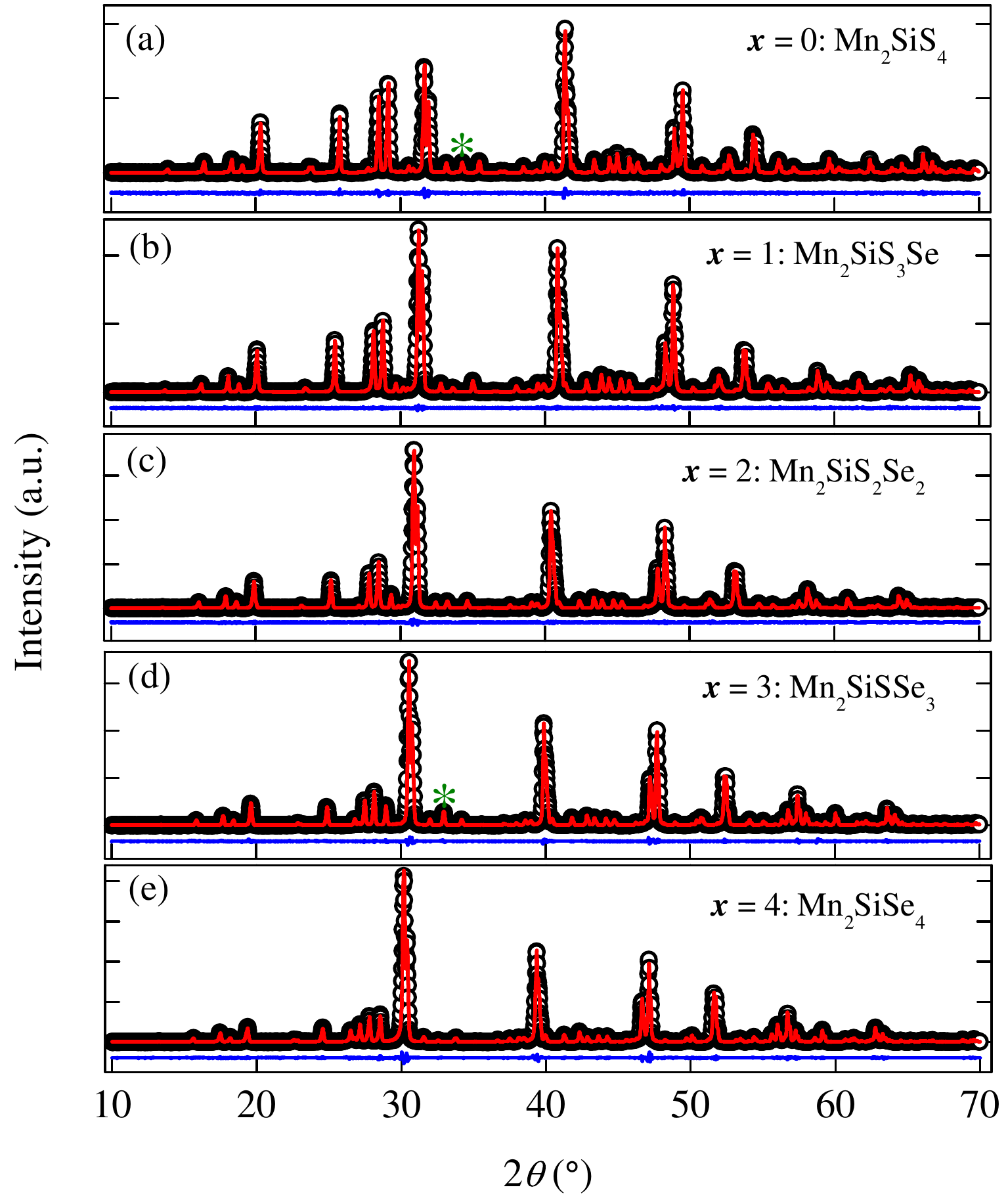}
	\caption{(color online) (a-e) The PXRD patterns of {\mssse} ($x$ = 0$\textendash$4) compositions are presented along with the results of structural analysis using $Pbnm$ space group. The black markers represent experimental data and the red lines are the fit using Le Bail approach. The blue horizontal line shows the difference curve. In (a) and (d), an asterisk marks a minor impurity phase, $\alpha$-MnS ($\approx$ 2wt$\%$).	 \label{fig_xrd}}
\end{figure}
\begin{table}[!t]
	\setlength{\tabcolsep}{5pt}
	\caption{The lattice parameters and the goodness-of-fit parameter ($R_\mathrm{wp}$) and the $\chi^2$ for the {\mssse} compounds at 300~K. All compositions crystallize in the orthorhombic $Pbnm$ space group. \label{tab1}}
	\begin{tabular}{lllllll} \hline \hline 
		& $a~({\AA})$ & $b$~({\AA}) & $c$~({\AA}) & $R_\mathrm{wp}$ & $\chi^2$ \\ \hline
		Mn$_2$SiS$_4$		&  12.692(9)  &  7.435(3) & 5.941(3) & 3.94$\%$ & 2.89 \\
		Mn$_2$SiS$_3$Se   &  12.860(1) &  7.527(4)  & 6.009(8) & 2.48 & 1.23 \\
		Mn$_2$SiS$_2$Se$_2$  & 13.000(3) & 7.605(6) & 6.076(5) & 2.31 & 1.42 \\ 
		Mn$_2$SiSSe$_3$   &  13.150(8) & 7.690(9) & 6.156(1)  & 1.78 & 1.35 \\
		Mn$_2$SiSe$_4$  & 13.302(8) & 7.777(2) & 6.243(6)  & 1.97 & 1.4 \\ \hline \hline
	\end{tabular}
\begin{tabular}{lllll}
	 & $d_\mathrm{Mn-S}$ {(\AA)} & $d_\mathrm{Si-S}$ {(\AA)} & $d_\mathrm{Mn-Mn}$ {(\AA)} \\ \hline 
	 	Mn$_2$SiS$_4$	& 2.5616(4) & 2.193(4) & 3.9112(3) $\times$ 2 \\
	 	                             & 2.6391(5) & 2.0375(3) & 3.7169(6) \\
	 	                             & 2.6226(5) &                  & 4.787(9) (along c) \\
	 Mn$_2$SiSe$_4$	& 2.7034(4) & 2.298(4) & 4.0997(5) $\times$ 2 \\
                            	 & 2.7513(1) & 2.3118(6) & 3.8876(4) \\
                            	 & 2.7403(9) &                  & 4.7153(6) (along c) \\	\hline                             
\end{tabular}
\end{table}

\subsection{Computational methods}
The density functional theory (DFT) 
calculations were carried out using the 
Vienna ab initio simulation package (VASP) \cite{kresse1993ab, furthmuller1994ab, kresse1996efficiency, kresse1996efficient}. 
The projector augmented wave (PAW) method 
was utilized for the electron-ion interaction
\cite{blochl1994projector, kresse1999ultrasoft, blochl2003projector} 
with an energy cutoff of 470~eV for the 
plane-wave basis functions. 
The generalized gradient approximation to 
exchange-correlation functional by Perdew, 
Burke, and Ernzhofer \cite{perdew1996generalized} 
was used. 
A $\Gamma$-centered (4$\times$4$\times$4) $k$-point 
grid based on Monkhorst-Pack scheme \cite{monkhorst1976special} 
was employed for initial structure optimization 
and later a finer grid of   6$\times$11$\times$13 
was used for further refinement. 
We have relaxed the structures until 
the Hellmann-Feynman forces on the ions were 
lower than 0.04~eV/{\AA}. 
An initial spin moment of 5~$\mu_B$ to Mn 
ions were assigned and the spin moment 
was allowed to relax. 
We also used VESTA \cite{momma_vesta_2011} 
software package 
for generating the crystal structures.

\section{Results and Discussion}
\subsection{X ray diffraction and air stability}
Powder X-ray diffraction (PXRD) patterns along with the 
Pawley fitting of {\mssse} ($x$ = 0$\textendash$4) compounds 
measured at room temperature 
are shown in Fig~\ref{fig_xrd}. The results of the structural analysis of
the PXRD patterns are summarized in Table~(\ref{tab1}). All samples 
crystallize in the orthorhombic $Pbnm$ space group ($\#$ 62). 
The compounds, Mn$_2$SiS$_3$Se ($x$ = 1), Mn$_2$SiS$_2$Se$_2$ ($x$ = 2), 
and Mn$_2$SiSe$_4$ ($x$ = 4) were obtained as 
pure phase samples, whereas the Mn$_2$SiS$_4$ ($x$ = 0) and  Mn$_2$SiSSe$_3$
($x$ = 3) samples contained minor impurity phase of 
$\alpha \textendash$MnS quantified to be less than 2~wt$\%$. 
\\
\indent
Air stability of \mssse\ is very important feature while considering 
use in practical device applications. 
Being a non-oxide, many chalcogenide based materials are prone 
to degradation upon exposure to air and moisture
\cite{choudhury_synthesis_2007,flanagan_enhanced_2015,
	de_kergommeaux_surface_2012}.
Air stability of {\mssse} compounds were investigated 
for over a period of 6 weeks by keeping the powder sample 
exposed to the ambient atmosphere. PXRD was 
collected regularly during this period and analyzed. 
PXRD patterns of the as-synthesized samples and 
those of the samples after exposure to air for 6 weeks 
showed no appreciable differences (not shown).
After 6-weeks exposure to air, 
no additional peaks or peak broadening was 
observed in any of the five compositions. We confirm 
that \msse\ series have good air stability thereby
establishing their potential for use in practical applications.
\\
\indent
The important structural feature of the \mssse\ compounds 
from the perspective of magnetism is the saw-tooth like
triangular arrangement of Mn lattice.\cite{hagemann_geometric_2000}
Such a lattice forms the basis for a frustrated lattice
depending on the different bond lengths or the exchange parameters
related to the triangular motif building up the saw tooth.
Mn has two crystallographically distinct positions in this 
structure, {\em viz.,} $4a$ and $4c$ where there are
four magnetic ions per cell with inversion symmetry and
mirror symmetry respectively. Previous neutron powder
diffraction studies on the $x$ = 0 compound in the temperature
range 4.2~K$\textendash$300~K have shown that there is no
structural change in the temperature range mentioned above.
For all the \mssse\ compounds, we assume the olivine structure
in the entire temperature range employed in the present study.
The refined lattice parameters that we 
obtain in the present study for \msse\
match well with the earlier report on the 
crystal structure.\cite{fuhrmann_structure_1989}
Incidentally, a structural peculiarity that the Mn(1)
octahedra being less distorted than the Mn(2) octahedra
was mentioned in Ref[\onlinecite{fuhrmann_structure_1989}]. 
Similarly the lattice parameters obtained for \msse\
series in the present work also matches well with the 
reported values.\cite{jobic_structure_1995}

\subsection{Specific heat}
\begin{figure}[!t]
	\centering
	\includegraphics[scale=0.34]{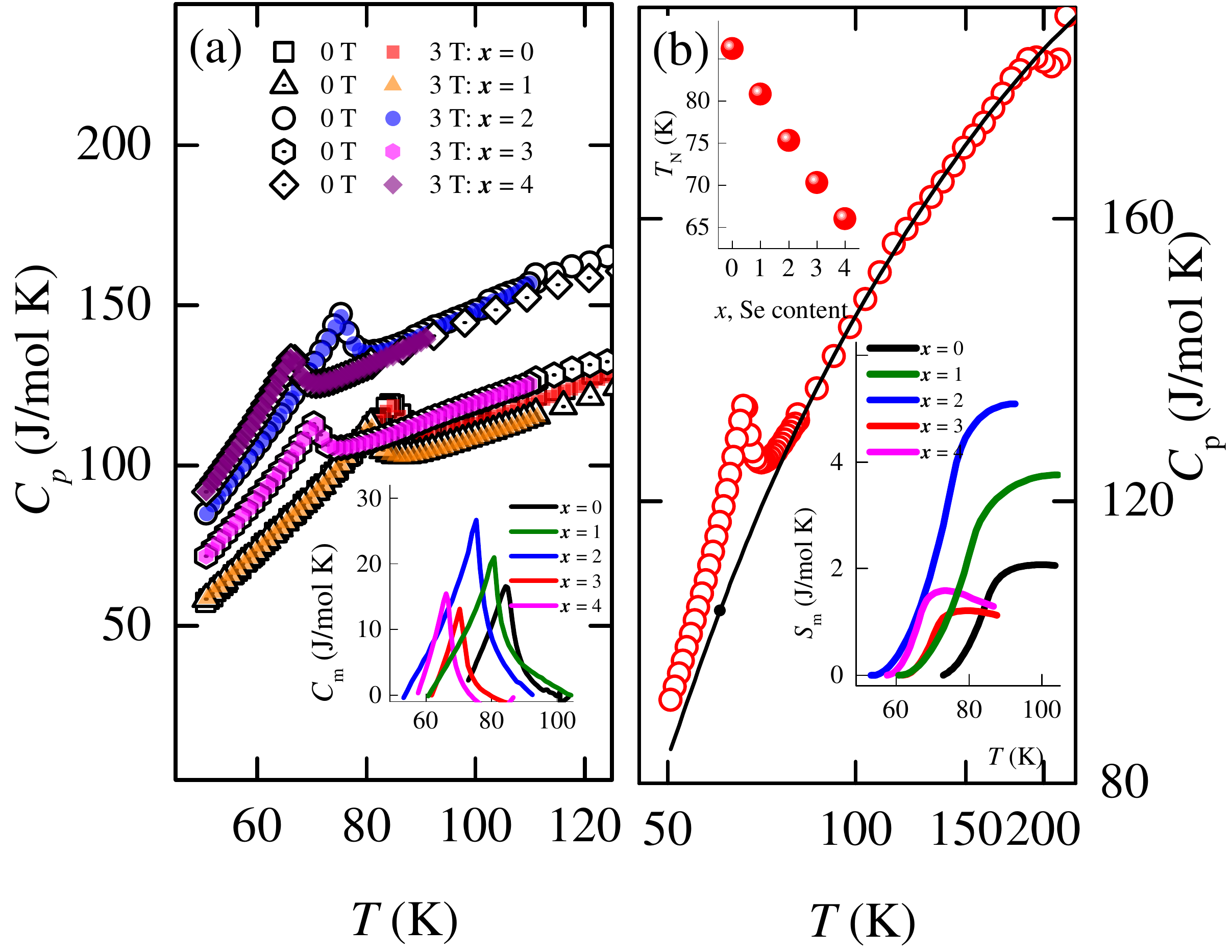}
	\caption{(color online)  (a) The specific heat of {\mssse} ($x$ = 0$\textendash$4) obtained in 0~T and 3~T are plotted together, showing no significant influence of the external magnetic field on the $T_N$s. (b) Shows the $C_p(T)$ of {\mss} (red circles) along with the curve fit using Einstein model (solid line). The inset of (a) shows the magnetic specific heat, $C_\mathrm{m}$. The top inset in (b) shows that, with increasing $x$, the $T_N$ decreases from 86~K to 66~K while the entropy, $S_\mathrm{m}$, is shown in the bottom inset. Significantly low value of entropy compared to $R$~ln$(2S + 1)$ is released at the magnetic transition in \mssse\ chalcogenides.\label{fig_cp}}
\end{figure}

\indent
The experimentally measured specific heat of 
{\mssse} ($x$ = 0$\textendash$4)
are presented in Fig~\ref{fig_cp} (a) where 
the specific heat under 0~T and
3~T are plotted together. The parent composition, 
{\mss} reproduces the antiferromagnetic
phase transition at $T_N \approx$ 84~K\cite{ohgushi_anomalous_2005,ohgushi_anomalous_2006,ohgushi_spin-flop_2007,junod_specific_1995}
which characterizes the paramagnetic-to-
antiferromagnetic phase transition.
It is reported that in the temperature range 
83~K$\textendash$86~K, {\mss}
displays WF; further, 
below 83~K it is an antiferromagnet.
From the present $C_p(T)$ data of {\mss}, we 
identify the AF transition at 86.2~K
by taking the derivative, $dC_p(T)/dT$. The 
WF transition reported at 83~K is less-conspicuous
in our derivative plot (not shown).
Under the application of 3~T magnetic field, 
no changes to the peak at $T_N$
is noticeable for any of the compositions 
$x$ = 0 to 4. 
This points towards strong AF nature of 
the underlying spin 
structure, up to atleast 3~T.
Upon substituting S with Se, the transition 
temperature $T_N$ decreases
from 86~K for $x$ = 0 to 66~K for $x$ = 4, 
{\msse}. The evolution of the
$T_N$ as a function of $x$ is presented in 
the inset (b) of Fig~\ref{fig_cp}.
\\
\indent
In order to account for the phonon part of the 
specific heat of \mssse, an Einstein model-based
curve fit was administered to the $C_p(T)$. 
Such a fit is demonstrated in
Fig~\ref{fig_cp} (b) for the case of {\mss}. 
In (b), the solid line represents the fit
using the following expression:
\begin{equation}
C_\mathrm{Einstein} = 3rR \sum_i a_i [x^2_ie^{x_i}/(e^{x_i} - 1)^2]
\end{equation}
where, $x_i$ = $\hbar \omega_\mathrm E/k_\mathrm B T$ and $a_i$ is the
weight factor for each mode.
The specific heat data in the temperature 
range, $T > 100~K$ was used for the fit.
We obtain the Einstein temperatures 
as $\theta_\mathrm{E1}$ = 744~K
and $\theta_\mathrm{E2}$ = 128~K.
The lattice part of the specific heat 
thus obtained was subtracted from
the total specific heat to obtain the 
magnetic part, $C_\mathrm{m}$ which
is plotted in the inset of Fig~\ref{fig_cp} (a) 
for all the compositions of {\mssse}.
The magnetic entropy, 
$S_\mathrm{m}$ = $\int (dC_\mathrm{m}/T)dT$
is calculated and plotted in the inset of 
Fig~\ref{fig_cp} (b)
for $x$ = 0$\textendash$4.
Though the {\mssse} compounds undergo a PM-AF 
second order phase transition,
it can be seen that significantly 
low magnetic entropy is released
at the $T_N$. The Mn$^{2+}$ with spin 
$S$ = 5/2 contributes
$R$~ln(6) = 14.8 J/mol-K towards spin entropy. 
In the case of {\mss}, only 14$\%$ of this value is released at the $T_N$.
This, in turn, suggests that the Mn$^{2+}$ spins of {\mssse}, which
form a two-dimensional saw-tooth-like triangular arrangement
are indeed in a frustrated magnetic state. 
Hence significant short-range magnetic order is expected to
coexist along with the prominent AF order.
It is noted here that the specific heat
analysis that we have performed is on the
data limited to only 50~K. Hence a comprehensive
estimation of the lattice specific heat
including a Debye term and extending down to
low temperature was not possible. 
This would have resulted in a deviation in the
values of $S_\mathrm{m}$ presented here.
However, we obtain supporting values from the
earlier reports on the specific heat analysis
and magnetic entropy determination of 
\mss.\cite{junod_specific_1995}
\subsection{Magnetization}
\begin{figure}[!b]
	\centering
	\includegraphics[scale=0.52]{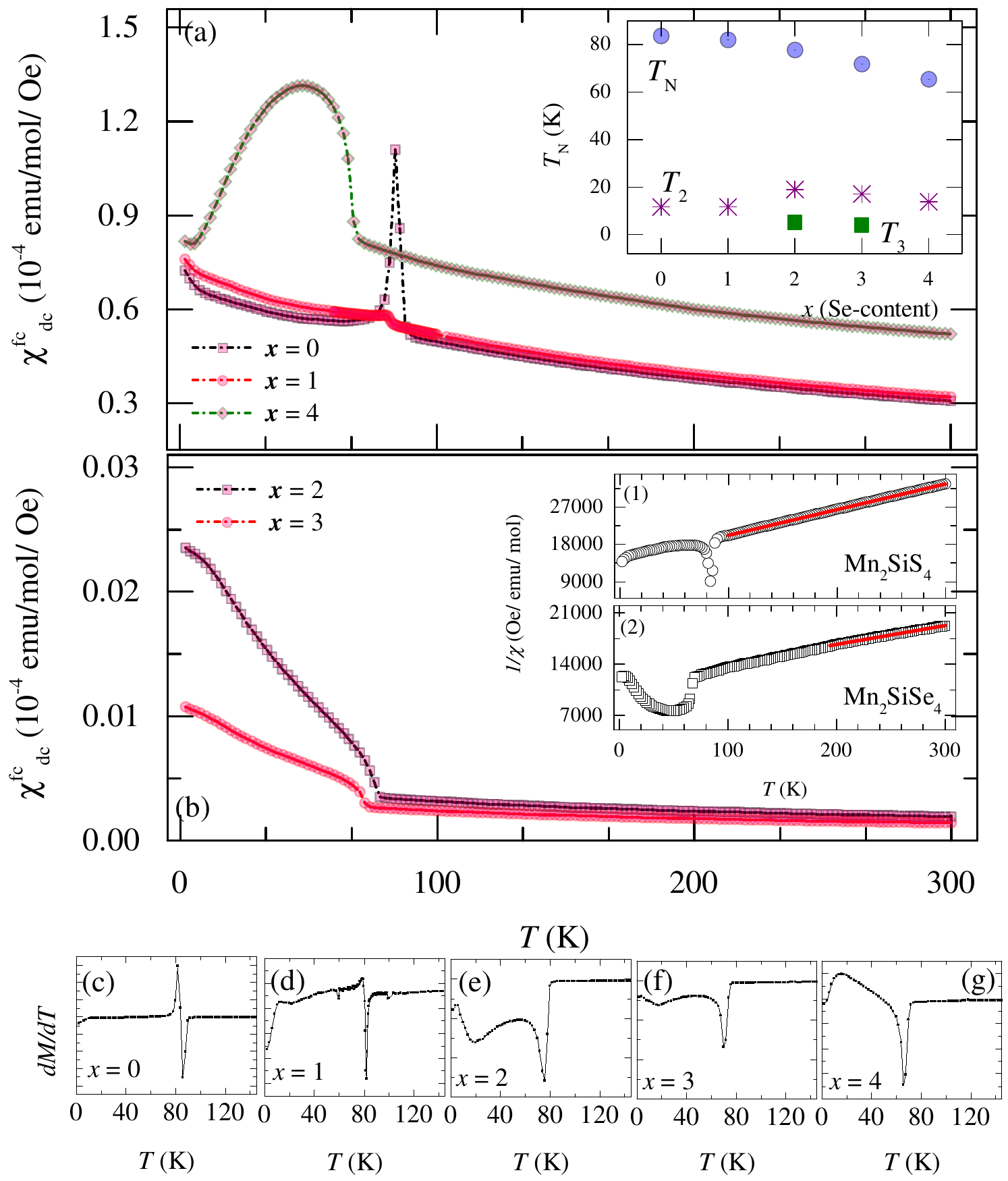}
	\caption{(color online)  The magnetic susceptibility of {\mssse} ($x$ = 0$\textendash$4) obtained in 500~Oe field cooled condition presented for the $x$ = 0, 1, 4 in panel (a) and $x$ = 2, 3 in panel (b). The inset of panel (a) shows the $T_N$'s as a function of Se-content ($x$). The insets (1) and (2) in panel (b) shows the 1/$\chi^\mathrm{fc}_\mathrm{dc}(T)$ curves of Mn$_2$SiS$_4$ and Mn$_2$SiSe$_4$ respectively along with Curie-Weiss fit (red solid line). (c-g) The derivative, $dM/dT$ versus temperature showing the multiple anomalies present in each composition.\label{fig_mag1}}
\end{figure}

\indent
The dc magnetic susceptibility, 
$\chi^\mathrm{fc}_\mathrm{dc}(T)$,
of the \msse\ series measured
in an applied magnetic field of
500~Oe are presented
in Fig~\ref{fig_mag1} for $x$ = 0, 1, 4
in panel (a) and $x$ = 2, 3 in (b). 
Though the phase transition temperatures ($T_N$)
identified in the specific heat data
are reflected in magnetic susceptibility as well,
a significant difference in the 
magnitude of magnetic susceptibility
is observed for the two sets of compositions in 
the panels (a) and (b). 
The magnetic phase transition in the 
case of \mss\ occurs as a sharp anomalous 
peak at $T_N$ = 83.7~K and matches with 
the reported value.
\cite{junod_specific_1995,lamarche_neutron_1994}
Upon progressive replacement of S with 
Se, the peak at the phase transition is 
weakened, and eventually for \msse\, a very
broad feature is seen below $\approx$ 65~K. 
This observation also matches with the 
previous report of the magnetic behaviour of
\msse.\cite{jobic_structure_1995}
\\
\begin{table*}[!t]
	\setlength{\tabcolsep}{8pt}
	\caption{The magnetic transition temperatures ($T_N$), effective paramagnetic moment ($\mu_\mathrm{eff}$), Curie-Weiss temperature ($\theta_{cw}$) of different compositions of the {\mssse} compounds. The $T_N$s are determined from the derivative of magnetic susceptibility. \label{tab2}}
	\begin{tabular}{llllllllll} \hline \hline 
		& $T_N$ & $T_2$ &  $T_3$ & $\mu_\mathrm{eff}$  & 	$\theta_{cw}$  & $f$ &  $T_N/\theta_{cw}$& $T_2/\theta_{cw}$ & $T_3/\theta_{cw}$\\ 
		& (K)   & (K)   & (K)   &  ($\mu_\mathrm{B}$/Mn) & (K) &  $|\theta_{cw}/T_N|$ & & & \\ \hline\hline
		Mn$_2$SiS$_4$		& 83.7 & 11.7 &     & 4.0(3) &  $\textendash$ 226  & 2.7 & 0.36 & 0.05 & \\
		Mn$_2$SiS$_3$Se    & 81.9 & 11.7  & & 4.07(2) &  $\textendash$ 221 & 2.7 & 0.36 & 0.05 & \\
		Mn$_2$SiS$_2$Se$_2$  & 77.7 & 19 & 5 &  3.95(5)  &  $\textendash$ 219 & 2.8 & 0.35 & 0.08 & 0.02 \\
		Mn$_2$SiSSe$_3$   &  71.7 & 17 & 3.8  & 3.8(2) &  $\textendash$ 193 & 2.7 & 0.36 & 0.08 & 0.02 \\
		Mn$_2$SiSe$_4$  & 65.5 & 13.7 & & 5.9(4) &  $\textendash$ 336  & 5.2  & 0.19 & 0.04 & \\ \hline \hline
	\end{tabular}
\end{table*}
\indent
The inset of (a) shows the $T_N$ values 
estimated from the
$\chi^\mathrm{fc}_\mathrm{dc}(T)$ data 
by taking the derivative, 
d$\chi^\mathrm{fc}_\mathrm{dc}(T)$/dT. 
The derivative, $dM/dT$ as a function 
of temperature for $x$ = 0$\textendash$4
are presented in the panels (c) to (g).
The magnetic transition temperatures, 
$T_N$s, thus estimated through the derivatives
are collected in Table~\ref{tab2}. 
From the magnetization data, we have 
been able to identify multiple
magnetic anomalies at low temperatures for 
all the compositions in \mssse.
For the $x$ = 0, 1 and 4 compositions,
in addition to the $T_N$, a low temperature
anomaly is observed in the temperature 
range near 12~K (denoted as $T_2$ in the table).
For the $x$ = 2 and 3 compositions, we observe
two more anomalies $T_2$ and $T_3$ in addition to 
the $T_N$.
This points out that the magnetic structure and
the low temperature magnetism of \mssse\ compounds
are more complex than the PM-AFM-WF transitions
that were reported earlier.
\cite{junod_specific_1995,jobic_structure_1995,ohgushi_anomalous_2005}
The presence of $\alpha$-MnS found in two of the
samples through x ray diffraction analysis does not
influence the magnetism as $\alpha$-MnS has a
magnetic transition at $T$ = 140~K but we do not
observe any anomalies at this temperature in
any of the compositions.
The effective paramagnetic moment, 
$\mu_\mathrm{eff}$ and the 
Curie-Weiss temperature, $\theta_{cw}$ 
are estimated form the 
inverse magnetic susceptibility versus 
temperature data following a curve-fit
to Curie-Weiss law. The insets (1) and (2) 
in panel (b) shows the 
representative Curie-Weiss fits administered on 
\mss\ and \msse\ as red solid lines.
The estimated parameters from the fit 
are collected in Table~\ref{tab2}
for all the five compositions.
Slightly diminished values of effective moment
compared to the theoretical spin-only moment of 
Mn$^{2+}$ in $d^5$ state,
$\mu_{th}$ = 5.92~$\mu_\mathrm{B}$, is 
observed in all the compounds
except for \msse. 
The Curie-Weiss temperature returns 
negative values which indicate that 
the overall magnetic interactions in these compounds
are antiferromagnetic type. 
The frustration parameter, 
$f$ = $\frac{|\theta_{cw}|}{T_N}$, shows a 
value of nearly 2.7 for all the 
compositions except for \msse\, for which
a higher value of 5.2 is recovered. 
This indicates that the compound
\msse\ is significantly frustrated than 
the other compounds. The frustration 
in \mssse\ stems from the geometrical 
triangular saw-tooth like arrangement 
of Mn$^{2+}$ spins.
The value of $f$ obtained for \mss\ 
compares well with the value reported
for this material earlier\cite{ohgushi_anomalous_2005}.
The magnetic frustration effect that is observed 
through the frustration index $f$
is supported by the structural 
feature of the triangular 
Mn arrangement (refer Fig~\ref{fig_str}) 
that makes up
the saw-tooth like lattice. In the 
case of \mss\, which has $f$ = 2.7, 
the Mn-triangle has two equal distances, 
3.97~{\AA} and one 3.71~{\AA}. 
However, in the case of \msse\, which 
has a high $f$ value (5.2), 
the Mn-triangle has all equal 
distance, 2.84~{\AA}.
\\
\indent
The magnetic structure of the 
$x$ = 0 and 4 compounds, \mss\ and \msse\ 
have been elucidated through neutron 
powder diffraction methods.
\cite{junod_specific_1995,lamarche_neutron_1994,bodenan_low-temperature_1996}
The neutron diffraction study of \mss\ 
confirmed the presence of olivine crystal structure
in the whole temperature range of 4.2~K to 180~K.
However, as noted previously, 
Mn in this structure occupies 
two distinct Wyckoff positions,
$4a$ and $4c$. At 4.2~K, the Mn moments 
in both the positions were
found to be collinear with the $y$-axis. 
At higher temperature, the magnetic moment
on the $4a$ site gradually rotates away from the $y$-axis. 
Between 83-86~K, both the $4a$ and $4c$ moments 
tend to align along the $x$-axis. However, 
the moments are subjected to canting in the $x$-$z$
plane.
\begin{figure*}[!t]
	\centering
	\includegraphics[scale=0.72]{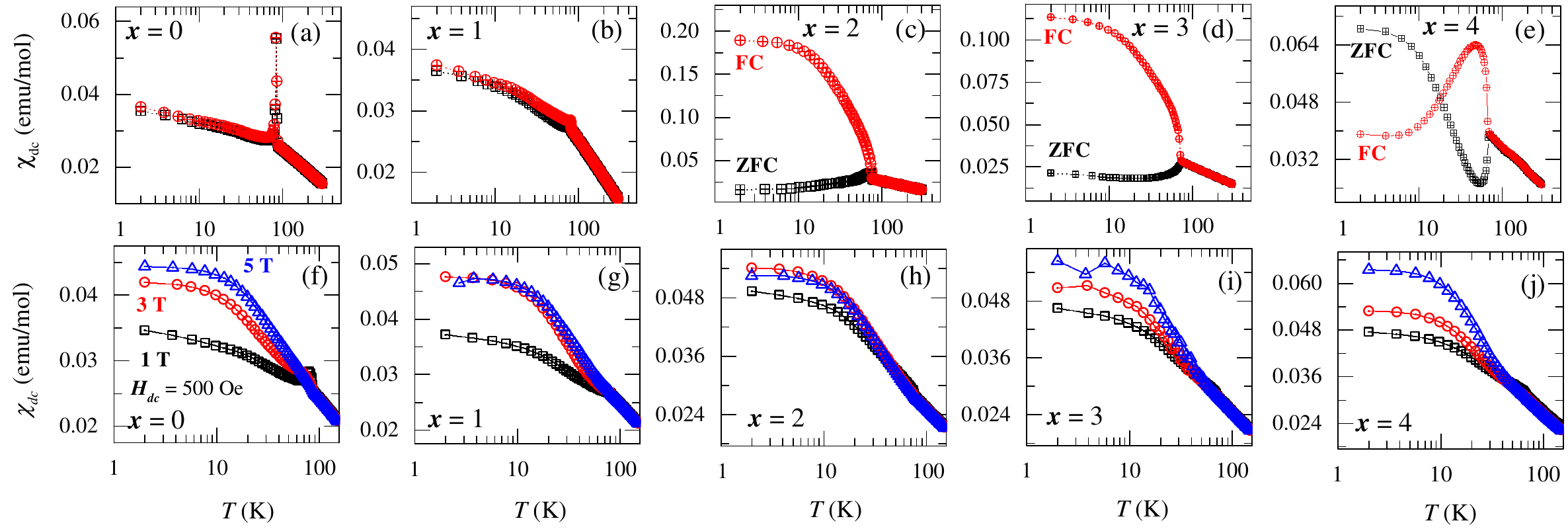}
	\caption{(color online)  (a-e) The dc magnetic susceptibility of {\mssse} ($x$ = 0$\textendash$4) obtained in ZFC and FC protocol using an external field of 500~Oe. Large bifurcation in the ZFC and FC arms observed for $x$ = 2, 3 whereas crossing of the ZFC and FC seen for $x$ = 4. (f-j) Shows the dc susceptibility in field-cooled mode for 1~T, 3~T and 5~T for $x$ = 0$\textendash$4. Even with the application of 5~T, no significant enhancement of magnetic moment is obtained.\label{fig_mag2}}
\end{figure*}
A similar case of tendency for canting 
of the spins was observed in
\msse\ as well. However, in this case 
clear signature of diffuse magnetic
scattering was observed, especially for 
the peaks at 2$\theta$ = 22$^{\circ}$
and 48$^{\circ}$\cite{bodenan_low-temperature_1996}.
The magnetization of \msse\ was then 
attributed to the short-range ferro or
ferrimagnetic arrangement of canted spins.
The spin correlations as a part of the 
diffuse magnetic scattering is seen to
persist up to 102~K.
The observation to diffuse magnetic scattering
in \msse\ from the previous studies support the
frustrated magnetism observed through a 
high frustration index, $f$, and also low
magnetic entropy released at the $T_N$.
\\
\indent
Multiple magnetic phase transitions at low temperature
were observed in Mn and Fe orthosilicate olivines.
\cite{kondo1966magnetic, santoro1966magnetic}
In \mssse\ compounds, more than one magnetic anomaly
is observed for all the compositions below 
their $T_N$ (refer to Table~\ref{tab2}).
By employing a Weiss mean-field model, the 
magnetic transitions in the Mn and Fe
orthosilicates were qualitatively understood 
based on the parameter $\gamma/\alpha$ which is the 
ratio of the two superexchange angles present in the
spin structure of these magnets. 
The ratio $\gamma/\alpha$ compares to
the ratio, $T_N/\theta_{cw}$. 
It was shown that in the special case
where $2 < \gamma/\alpha < 1$, there arises a new low temperature
phase transition below $T_N$ and it is indicated 
by the low values of the ratio $T_2/\theta_{cw}$ 
where $T_2 < T_N$ is the low temperature transition.
In Table~\ref{tab2}, we have collected the ratios 
$T_N/\theta_{cw}$, $T_2/\theta_{cw}$
and $T_3/\theta_{cw}$ for the \mssse\ compositions.
It is easily noted that the value of $T_N/\theta_{cw}$ 
is relatively constant across the compositions, 
except for the highly frustrated composition, \msse.
Also, the values of $T_2/\theta_{cw}$ and $T_3/\theta_{cw}$ are 
highly diminished compared to that of $T_N/\theta_{cw}$. 
This is in agreement with the simple Weiss-field 
approach where the calculated exchange energies 
supported the low temperature magnetic anomalies. 
The broadness of the magnetic anomalies below
$T_N$ in the \msse\ compositions point toward short-range
magnetic order rather than a long-range magnetic order in to
a new magnetic structure. 
The dc magnetic susceptibility of the \mssse\ series 
in zero field-cooled (ZFC) and field-cooled (FC) protocol
in the presence of external magnetic field, $H_\mathrm{app}$
= 500~Oe is presented in Fig~\ref{fig_mag2} (a-e).
The effect of an external magnetic fields 1~T, 3~T and 5~T
upon the field-cooled magnetic susceptibility is shown
in the panels (f-j) of the same figure.
For the compositions $x$ = 0, 1, the ZFC and the FC
arms show no bifurcation at all (a, b). The magnetic phase
transition is evident as a strong anomaly thereby confirming
the AFM transition.
For the compositions $x$ = 2, 3, and 4 strong
irreversibilities are observed in the magnetic response
which is an indication of significant short-range magnetic
correlations or spin glass-like features.
Interestingly, the highly frustrated compound \msse\
presents a ZFC/FC response where the ZFC and FC
arms cross each other in the low temperature region.
This crossing happens at $T\approx$ 18~K.
This feature resembles the case of negative
magnetization observed in many other oxide systems.\cite{kumar2015phenomenon,
	bartolome2002negative,menyuk1960magnetization}
In the case of the spinel compound Co$_2$VO$_4$,
the negative magnetization was explained in terms
of a ferrimagnetic structure ($T_c$ = 158~K) and 
the resulting complex magnetism at low temperature.
However, the presence of ferromagnetic clusters
embedded in an AFM matrix also can display negative
magnetization as evidenced in the case of the rare-earth
manganite, NdMnO$_{3\pm\delta}$ where the stoichiometry
of the oxygens also seem to play a role.\cite{bartolome2002negative}
In order to accurately determine the presence of ferromagnetic
short-range ordered clusters, low temperature neutron diffraction
experiments were undertaken.
In the case of \msse, the magnetic structure that is proposed
already point towards the presence of short-range
ferro or ferrimagnetic arrangement of canted spins.
The magnetization data presented in Fig~\ref{fig_mag2} (e)
supports the claim of short-range ferrimagnetic
canted spins. However, a detailed low temperature
neutron powder diffraction study can shed more light
on the proposed magnetic features.
It can be noted that the application of external magnetic
fields up to 5~T does not produce appreciable enhancement of
the magnetization in any of the \mssse 
compositions (panels (f-j)).
\\
\begin{figure}[!b]
	\centering
	\includegraphics[scale=0.65]{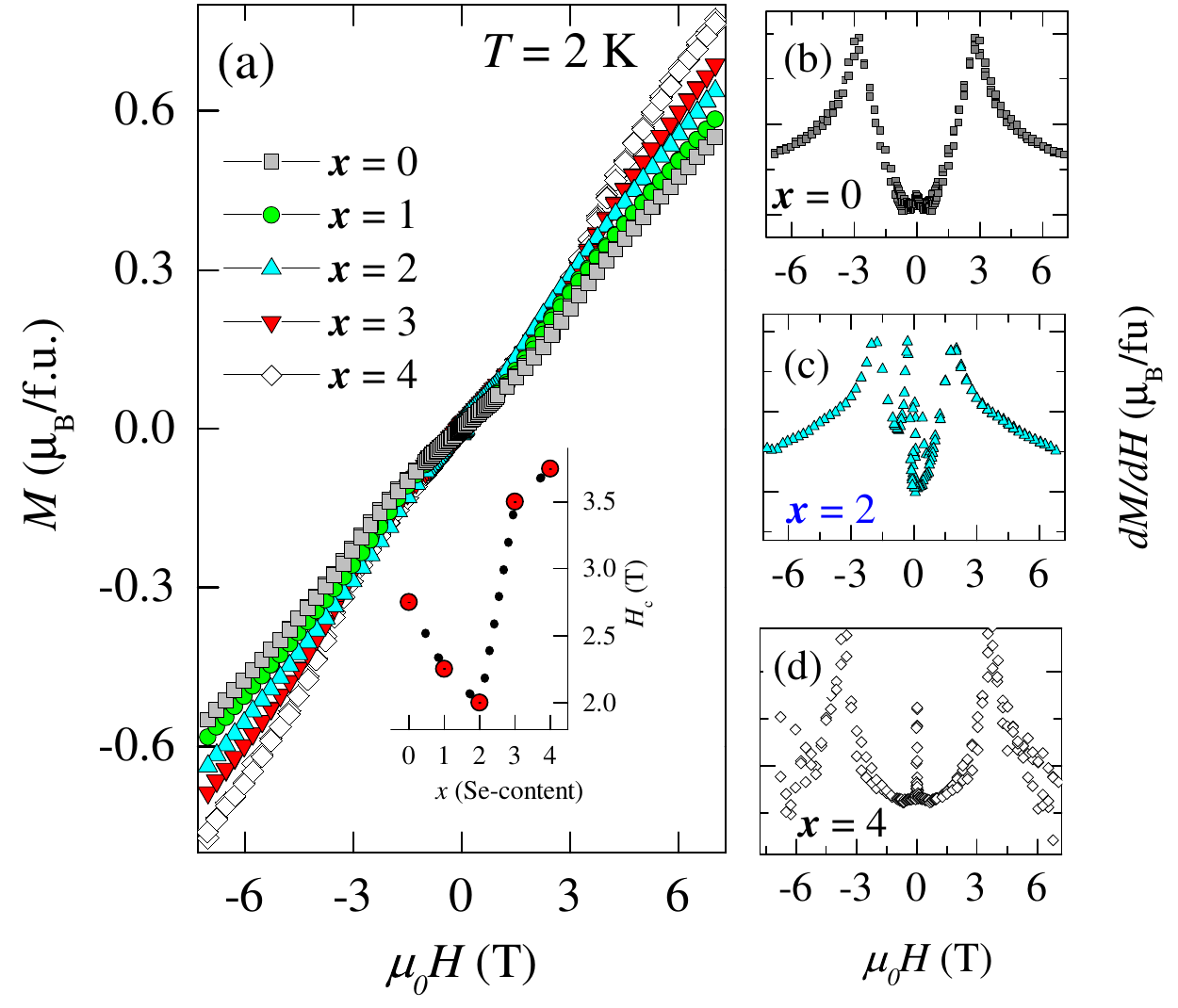}
	\caption{(color online) (a) The isothermal magnetization of {\mssse} ($x$ = 0$\textendash$4) at 2~K. The inset shows the evolution of the critical field, $H_c$, as  a function of Se content, $x$. (b-d) The derivative, $dM/dH$ versus temperature showing the presence of metamagnetic phase transitions.  \label{fig_mag3}}
\end{figure}
\indent
The isothermal magnetization curves, 
$M(H)$, as a function of applied field $H$
in the range $\textendash$7~T to +7~T at 2~K for 
\mssse\ $x$ = 0 $\textendash$ 4 are
plotted together in Fig~\ref{fig_mag3} (a).
The curves represent typical antiferromagnetic 
response with no indication of magnetic hysteresis. 
At 2~K and 7~T, the maximum magnetic moment attained
is about 0.77~$\mu_\mathrm{B}/$f.u. for \msse\. 
The lowest moment is attained 
for \mss\ which has a value of 0.55~$\mu_\mathrm{B}/$f.u.
One of the earliest work on the Mn-chalcogenide olivines
was related to the spin-flop transition and the associated
tricritical point in the $H \textendash T$ phase diagram.
\cite{ohgushi_anomalous_2005}
The field-induced spin-flop transitions are reproduced in 
all the compositions in the present series of \msse\ 
($x$ = 0$\textendash$4).
The field-induced spin-flop transitions
are clearly evidenced in the derivatives $dM/dH$ 
versus $H (T)$ plotted in Fig~\ref{fig_mag3} (b-d)
which are shown for $x$ = 0, 2 and 4. 
The variation of the critical field $H_c$ for spin-flop
as a function of the Se-content is presented in the inset of (a).
The spin-flop transition in \mss\ single crystals 
along the crystallographic $c$-axis
is observed at a critical field of about 3~T. 
In the present case, our samples of
polycrystalline \mss\ also display the spin-flop transition 
at a comparable field value of 2.7~T. 
It can be observed that the $H_c$ first decreases with 
the replacement of S with Se, until $x$ = 2.
Beyond $x$ = 2, for $x$ = 3, 4 $H_c$ increases and 
reaches a maximum for \msse\ which is in fact, 
the highly frustrated composition in this group.
In order to correlate the crystal and the electronic
structure of the \mssse\ and to explore the band structure
peculiarities of the current compositions as compared
to that observed in Fe-based olivines that are predicted
thermoelectrics \cite{gudelli_predicted_2015}, we now take
a look at the results from density functional theory calculations.

\subsection{Density functional theory and thermal conductivity}
\begin{figure}[!t]
	\centering
	\includegraphics[scale=0.35]{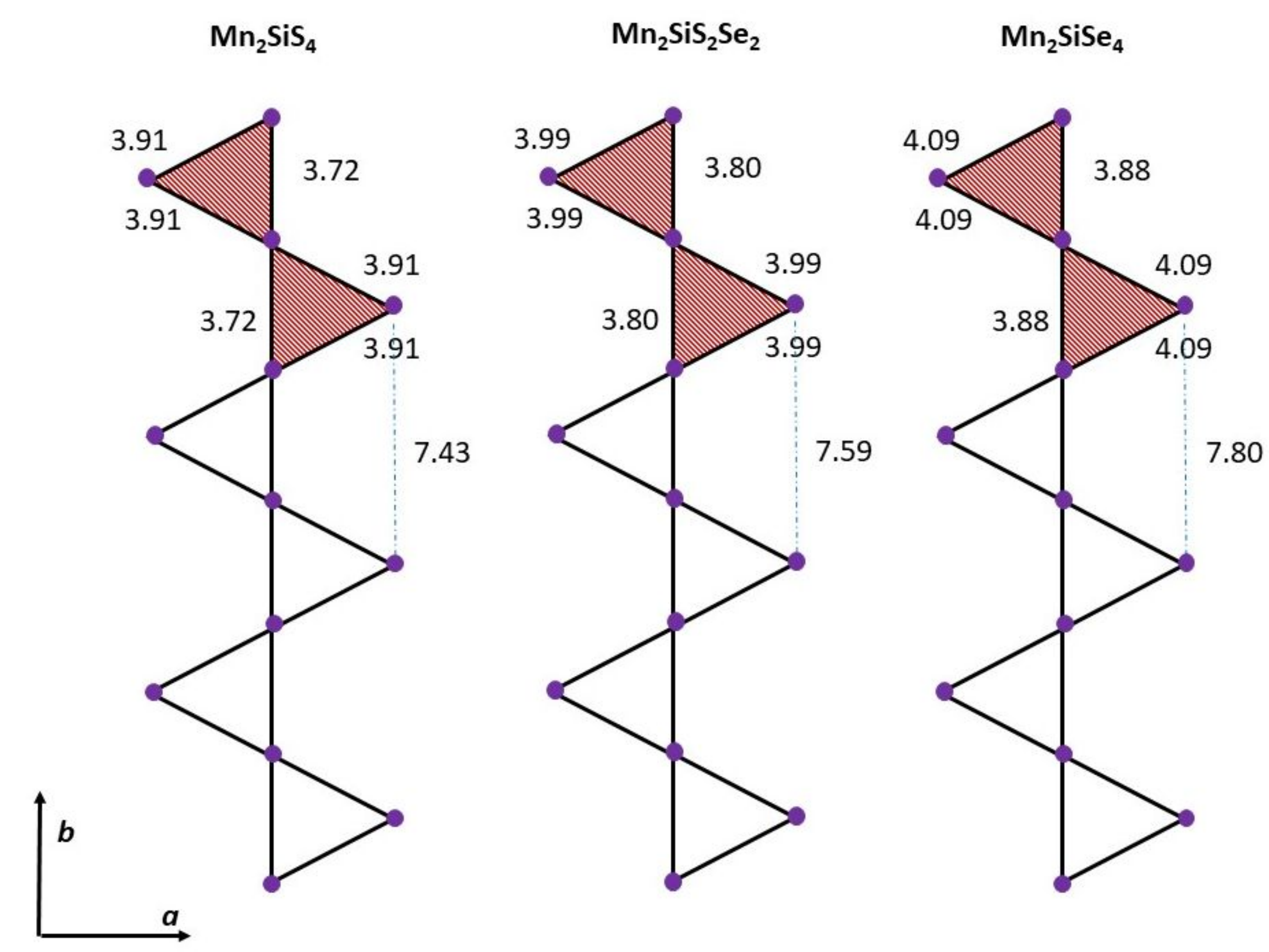}
	\caption{\label{fig_sawtooth} (color online) 
		A schematic figure showing the saw-tooth chain of Mn in \mss, Mn$_2$SiS$_2$Se$_2$ and \msse. The distance between Mn in m$_1$ and m$_2$ positions are marked.
	The Mn-triangles tend to form isosceles in all the \mssse\ compositions. The
distance between the 2-triangle units (shaded region) along $b$-direction increases as the Se-content increases. The interlayer distances between the saw-tooth layers (in $ab$-plane) also increases with higher Se-content.}
\end{figure}

The magnetic structure of \mss\ and \msse\ are
reported in the antiferromagnetic structure through
neutron diffraction studies \cite{bodenan_low-temperature_1996, lamarche_neutron_1994}.
\begin{table*}
	\setlength{\tabcolsep}{2pt}
	\caption{\label{tab:dft} The total energy ($E_\mathrm{AFM}$), theoretical and experimental lattice parameters $( (a, b, c)_\mathrm{th}$ and $ (a, b, c)_\mathrm{exp} )$ magnetic moments ($\mu_{t}$/Mn and $\mu_{c}$/Mn) and the calculated band gaps from the DFT calculations.}
	\begin{tabular}{llllccc} \hline\hline
		Composition	& $E_\mathrm{AFM}$ (eV) & $ (a, b, c)_\mathrm{th}$ ({\AA}) & $ (a, b, c)_\mathrm{exp}$ ({\AA}) & $\mu_{c}$/Mn ($\mu_B$) & $\mu_t$/Mn ($\mu_B$) & $\Delta_b$ (eV) \\ \hline\hline
		\mss\ & 0.0 & 12.46, 7.27, 5.87 & 12.69, 7.44, 5.94 & 4.04, -4.05 & 4.0 & 0.47 \\
		& 0.40 & 12.69, 7.44, 5.94 & & 4.12, -4.16 & & 0.64 \\ \hline
		Mn$_2$SiS$_3$Se (a) & 0.34 & 12.726, 7.320, 5.892 & 12.860, 7.527, 6.009 & 4.03, -4.03 & 4.07 & \\ 
		Mn$_2$SiS$_3$Se (b) & 0.35 & 12.530, 7.315, 6.013 &  & 4.04, -4.04 & & 0.40 \\
		Mn$_2$SiS$_3$Se (1) & 0.0 & 12.59, 7.35, 5.91 & & 4.02, -4.02 &  & 0.32 \\
		Mn$_2$SiS$_3$Se (2) & 0.02 & 12.583, 7.335, 5.942 & & 4.02, -4.03 &  & \\
		Mn$_2$SiS$_3$Se (3) & 0.12 & 12.590, 7.34, 5.96 & & 4.04, -4.02 &  & \\
		Mn$_2$SiS$_3$Se (4) & 0.10 & 12.603, 7.338, 5.915 & & 4.02, -4.03 &  & \\ \hline
		Mn$_2$SiS$_2$Se$_2$ & 0.0 & 12.538,  7.432,  5.893 & 13.00, 7.61, 6.08 & 3.97, -3.97 & 3.95 & \\
		Mn$_2$SiSSe$_3$ & 0.0 &  12.75,  7.45,  5.92 & 13.150, 7.690, 6.156 & 3.92, -3.92 & 3.8 &  0.1\\
		\msse\ &  0.0 &  &  13.30,  7.78,  6.24 & 4.12, -4.12 & 5.9 & 0.45\\ \hline\hline 
	\end{tabular}
\end{table*}
The magnetic moments of Mn are proposed to lie
in the $c$ direction; the spins in the $ab$ plane
are ferromagnetically coupled while the adjacent layers
are coupled antiferromagnetic.
We have performed the DFT calculations for the 
antiferromagnetic spin arrangement of the 
four different compositions in \mssse.  
The experimental crystal structure for 
\mss, obtained from the room temperature
structure determined in the present work, 
was used as the starting point to generate the 
initial structures for all the four compounds
following a energy-minimization process. 
The crystal symmetry is found to be orthorhombic 
with six inequivalent positions, agreeing with the
$Pbnm$ space group reported \cite{fuhrmann_structure_1989}. 
The optimized lattice constants, spin magnetic moments, 
and the band gaps of these systems are summarized 
in Table~\ref{tab:dft} and presented along with the
corresponding experimental values for easy comparison. 
The total energy with the optimized lattice constants 
is lower than that with experimental lattice constants 
by 0.4~eV. 
The calculated magnetic moments on the Mn 
atoms show antiferromagnetic spin ordering.
The DFT-calculated magnetic structure was found
to match the experimental structure reported
through previous neutron diffraction experiments \cite{lamarche_neutron_1994}.
The range of the calculated spin magnetic moments, $\mu_{c}$, 
are presented in the table for the 
optimized structure. 
An agreement is found between the $\mu_{c}$ and 
the experimental values ($\mu_{t}$) obtained from magnetization
results of the present work. 
The substitution of one S by 
Se leads to two structures with the formula 
unit of Mn$_2$SiS$_3$Se which are identified as 
structures (a) and (b) in Table~\ref{tab:dft}. 
Similar substitution by two Se atoms also leads to 
two structures using the same set of 
inequivalent atom positions. 
Apart from these, we also tested mixed 
structures with random substitutional 
positions in a supercell which led to 
Mn$_2$SiS$_3$Se and Mn$_2$SiS$_2$Se$_2$ based on atom 
count only. However, we find that some of these 
structures have lower energy than the 
structures that conform to the 
orthorhombic symmetry and therefore 
we include them in the results,
identified as structures Mn$_2$SiS$_3$Se (1)-(4) 
in the Table~\ref{tab:dft}. 
For the Mn$_2$SiS$_2$Se$_2$ compound, 
substitution of inequivalent atoms 
leads to the lowest energy structure compared 
to random substitution and therefore we present 
only the lowest energy structure in the table. 
However, for this compound and for Mn$_2$SiSSe$_3$, random 
substitution can lead to low lying structures 
within 0.02$\textendash$0.07~eV above the ground state. 
\begin{figure}[!b]
	\centering
	\includegraphics[scale=1.8]{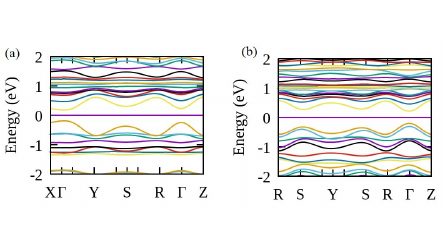}
		\includegraphics[scale=1.6]{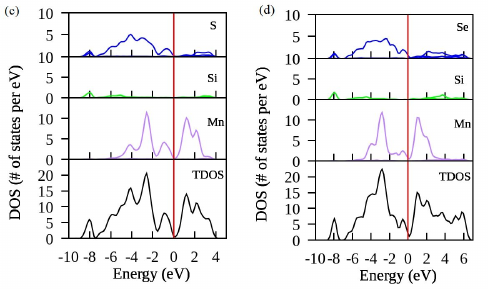}
		\includegraphics[scale=0.55]{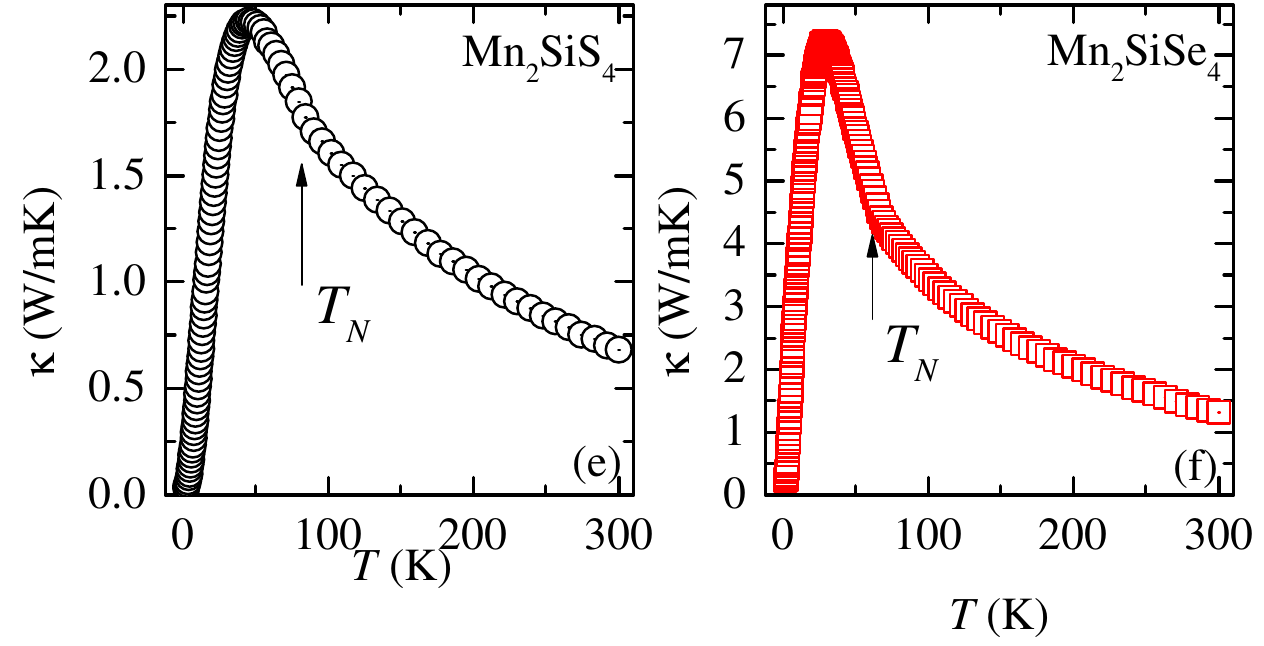}
	\caption{\label{fig_dft} (color online) 
		The DFT predicted band structure of (a) \mss\ and (b) \msse. 
		The electronic projected density of states 
		(PDOS) for \mss\ and \msse\ are shown in 
		(c) and (d) respectively. 
	     The experimentally measured thermal conductivity of
     (e) \mss\ and (f) \msse. The arrows mark the temperature of 
     antiferromagnetic phase transition.}
\end{figure}
The \msse\ compound with experimental lattice 
parameters is the lowest energy structure.
For all the five compounds studied, the 
antiferromagnetic phase is the stable magnetic phase. 
The trend of the spin moment of the mixed 
compounds compare excellently with the trend 
seen from experiment. 
With higher number of Se atoms in the mixed 
compound, the spin moment decreases. 
However, the spin moment of the 
\msse\ from DFT is much smaller compared 
to that derived from experimental data. 
We find that there can be several local minima 
in the potential energy surface with slightly 
different lattice constants for this compound. 
In all cases however the spin moment is still 
smaller than the experimental value. 
The experimental spin moments are determined 
from magnetization data collected at elevated temperatures 
where lattice expansion can lead to reduced 
interactions between atoms. 
The DFT calculations are done without any temperature 
effect which may explain the difference 
between the DFT and experimental spin moments. 
Moreover, these calculations also do not take 
into account non-collinear spin moments.
\\
\indent
The band structure of the two terminal 
compounds \mss\ and \msse\ are shown 
in Fig~\ref{fig_dft} (a, b) respectively. 
The bandgap of \mss\ and \msse\ was found to be 
0.47~eV and 0.45~eV respectively. 
Both structures have a direct 
bandgap as indicated by the valence band 
maximum (VBM) and conduction band maximum 
(CBM) from the band structure plot. The band 
structures have flat bands from G-X 
crystallographic direction and similar to 
that of Gudelli {\em et al.} \cite{gudelli_predicted_2015} 
in their Fe$_2$GeS$_4$ and Fe$_2$GeSe$_4$ band structure plots. 
The corresponding total and projected density of 
states (DOS) for \mss\ and \msse\ are 
represented in panels (c, d) of Fig~\ref{fig_dft}. 
The DOS shows that the states near the Fermi level 
arise mainly from the Mn d-states and S/Se $p$ states. 
In both the system the Si states lie deeper in energy.  
The conduction band has contribution mainly from the Mn $d$ states.
The band gap obtained in the present study
differ from those reported for \mss \cite{davydova2018thio}.
However, the antiferromagnetic spin arrangement 
assumed in the work by Davydova {\em et al}
seem to be different from the AFM structure
that is obtained in the present work as well as
in earlier neutron reports \cite{lamarche_neutron_1994}.
An LDA + U approach was used by Davydova {\em et al} which
influences the band gap since U can be adjusted to match the
experimental band gap.\\
\indent
Motivated by the band structure features that
we found in \mss\ and \msse\ and from the reports
on other thio-olivines that project these materials
as potential candidates for thermoelectric applications,
we measured the thermal conductivity, $\kappa (T)$.
Figure~\ref{fig_dft} (e, f) shows the $\kappa (T)$ for
\mss\ and \msse\ respectively.
Both compounds show thermal conductivity 
that is reminiscent of semiconducting 
materials where phonons dominate the thermal 
transport \cite{tritt2005thermal}. 
At low temperatures the thermal conductivity 
rapidly increases with increasing temperature, 
forms a pronounced maximum centered below 50~K, 
and then decreases down to room temperature. 
The maximum in low temperatures occurs due to 
reduction of the thermal scattering at low 
temperatures, {i.e.,} in the regime where the 
phonon mean free path becomes larger than 
the interatomic distances. For both compounds 
the magnetic phase transition appears only 
as a small kink in $\kappa (T)$ at the N\'{e}el 
temperature (marked by arrows in 
Fig~\ref{fig_dft} (e, f)). 
The band gap obtained for \msse\ is 
slightly diminished compared to that 
os \mss\ and the former shows $p$-type 
conductivity according to our Seebeck 
coefficient measurement (not shown).
\\
\indent
As the Se-content is increased from $x$ = 0 to $x$ = 4, 
the sharp magnetic transition observed in \mss\ is replaced
with a broad transition extending over a large temperature range
below 65~K. The results of magnetization and the specific heat experiments
point towards the emergence of short-range order stemming from
enhanced magnetic frustration.
In Fig~\ref{fig_sawtooth} the Mn-Mn distances between the
Mn atoms occupying the crystallographically distinct m$_1$ and m$_2$
positions are represented for \mss, Mn$_2$SiS$_2$Se$_2$ and \msse.
The distances marked on the figure are obtained from the refined
x-ray diffraction data which is cross-checked against the distances
obtained from DFT.
It can be seen from the figure that the Mn-triangles that form
the saw-tooth lattice are isosceles and the Mn-Mn distances between
the atoms in the m$_1$ chain (along $b$) increases with Se-content.
In general, the distances in the Mn-triangle increases from \mss\ towards \msse. The Mn(m$_1$)$\textendash$Mn(m$_1$) distance along the $c$-direction for \mss\ is 5.94~{\AA} which increases to 6.24~{\AA} for \msse. The Mn(m$_1$)$\textendash$Mn(m$_1$) distances
also undergo a similar increase. From this, it is clear that the
inter-layer distance between the saw-tooth layers increase towards
\msse\ and subsequently a weakening of the exchange interaction
can result. These structural features associated with the
saw-tooth triangles lead to the formation of spin clusters
in \mssse\ with increased Se-content resulting in predominant
short-range order.
\\

\section{Conclusions}
The magnetism of chalcogenide olivine \mssse\ with a saw-tooth lattice
for the Mn moments are studied in detail using magnetization, specific heat and first-principles density functional theory calculations.
Progressive substitution of S using Se in \mss\ is seen to shift the
antiferromagnetic transition temperature from 86~K to 66~K.
Though an antiferromagnetic transition is clear, the magnetic
entropy estimated from the analysis of specific heat reveals
diminished values suggesting strong spin fluctuations present.
Among the \mssse\ compositions studied, \msse\ is the most frustrated.
A non-linear trend in the evolution of the critical field
for spin-flop is found across the compositions.
Density functional theory calculations support the stable
orthorhombic crystal structure across the series and confirms
the antiferromagnetic structure for \mss\ and \msse.
Quasi-flat-band features similar to that seen in Fe-based olivines
are seen in the present case however, the experimental thermal transport results do not support features favourable for a good thermoelectric.

\section{Acknowledgements}
HSN acknowledges the UTEP start-up fund and UT Rising-STAR 
in supporting this work. BS acknowledges the financial support for this work 
provided by the University of Oklahoma startup funds. DA and KG acknowledge support from DOE’s (Basic Energy Science) Early Career Research Program.
RRZ and TB acknowledge support by Department of Energy Basic Energy Science through Grant Nos. DE‐SC0002168 and DE‐SC0006818. Support for computational time  by the NSF's XSEDE project through grant TG‐DMR090071 is gracefully acknowledged. 
AC and SRS acknowledge The University of Texas at
El Paso (UTEP) Start-up fund and NSF-PREM Program (DMR—1205302). 
The part of the paper prepared by Dr. Singamaneni and co-author A. Cosio are funded under the Award No. 31310018M0019 from UTEP, Nuclear Regulatory Commission. The statements, findings, conclusions, and recommendations are those of the author(s) and do not necessarily reflect the view of the UTEP or The US Nuclear Regulatory Commission.
%
%

 \end{document}